\documentclass[conference]{IEEEtran} 
\usepackage{cite}      % Written by Donald Arseneau
\usepackage{graphicx}  % Written by David Carlisle and Sebastian Rahtz
\usepackage{epsfig}
\usepackage{enumerate}
\usepackage{color}
\usepackage{subfigure} % Written by Steven Douglas Cochran

\usepackage{amsmath}   % From the American Mathematical Society

\begin{document}
% paper title
% \title{PLL Figure-of-Merit considering lock time}
\title{Tradeoffs between Settling Time and Jitter in\\ Phase Locked Loops}
% author names and affiliations
% use a multiple column layout for up to three different
% affiliations
% \author{\authorblockN{Pallavi Paliwal}
% \authorblockA{Dept. of Electrical Engg.\\
% Indian Institute of Technology\\
% Atlanta, Georgia 30332--0250\\
% Email: pallavi.paliwal@iitb.ac.in}
% \and
% \authorblockN{Homer Simpson}
% \authorblockA{Twentieth Century Fox\\
% Springfield, USA\\
% Email: homer@thesimpsons.com}
% \and
% \authorblockN{Shalabh Gupta}
% \authorblockA{Dept. of Electrical Engg.\\
% Indian Institute of Technology\\
% }}
% 
% avoiding spaces at the end of the author lines is not a problem with
% conference papers because we don't use \thanks or \IEEEmembership
% for over three affiliations, or if they all won't fit within the width
% of the page, use this alternative format:
% 
\author{\authorblockN{Pallavi Paliwal,
Mohanrao Sattineni,
Shalabh Gupta
\authorblockA{Department of Electrical Engineering, IIT Bombay, Mumbai – 400076, India\\
Email: pallavi.paliwal@iitb.ac.in, mohanrao.venkat@gmail.com, shalabh@ee.iitb.ac.in}
}}
% use only for invited papers
%\specialpapernotice{(Invited Paper)}
% make the title area
\maketitle
\begin{abstract}
In most PLL architectures, trade-off exists between settling time and jitter performance, which is ignored during Figure of Merit calculation. This work derives a new Figure of Merit for PLL, which has settling time as added performance parameter, along with jitter and power. Here, the trade-off between settling time and jitter is analyzed theoretically, and with behavioral simulations for (i) linear Time-to-Digital based PLL (ii) non-linear Bang-Bang Phase Detector based PLL (iii) Hybrid PLL with adaptive gain, to obtain settling time vs. jitter relation, based on which commonly used Figure of Merit is modified. To understand trade-off relation between settling time and jitter for Bang-Bang Phase Detector based PLL, this work also derives settling time equation for non-linear PLL, by using recursive time-domain equations.
\end{abstract}
% no keywords
% For peer review papers, you can put extra information on the cover
% page as needed:
% \begin{center} \bfseries EDICS Category: 3-BBND \end{center}
%
% for peerreview papers, inserts a page break and creates the second title.
% Will be ignored for other modes.
\IEEEpeerreviewmaketitle
\section{Introduction}
% no \PARstart
Phase Locked Loops, used in various applications, are required to generate frequency with low
power consumption, within low settling time and reduced spurious tones. But, in most PLL architectures, the trade-off existing between jitter, lock time and power does not allow all PLL performance parameters to be optimized simultaneously.

Figure of Merit (FoM) defined in \cite{ref:fom_paper}, which is commonly used to benchmark PLL performance, does not include effect of increase in settling time, while reducing loop gain to reduce jitter, as shown in Eqn. (\ref{eqn:fom}).
\begin{equation}
\label{eqn:fom}
FoM = 10log\left[\left(\frac{\sigma_t}{1s}\right)^2\left(\frac{P}{1mW}\right)\right]
\end{equation}

For example, PLL design in \cite{ref:7cicc_10}, having low jitter and power values,  has good Figure of Merit as per Eqn. (\ref{eqn:fom}). But, this architecture achieves low jitter (0.4ps) and low power (2.8mW), at the cost of very high settling time (300$\mu$s). This instance signifies that without considering settling time as a performance parameter, the Figure of Merit in Eqn. (\ref{eqn:fom}) is inadequate for benchmarking PLL performance.

In this regard, the trade-off between lock time and jitter is analyzed in this work, for different PLL architectures. Accordingly, commonly used Figure of Merit for PLL is modified, to benchmark PLL performance with consideration to all important specifications i.e. lock time, power and jitter.

This work is organized as follows. Section \ref{sec:bbpll}-\ref{sec:hybrid_pll} analyzes trade-off between settling time and jitter for  non-linear Bang-Bang Phase Detector based PLL (BBPLL), Time-to-Digital Converter (TDC) based linear PLL, and Hybrid PLL with adaptive gain. Theroretical analysis for trade-off inherent in these PLL systems, is backed up by results from behavioral simulation of  Verilog-A model of PLL architectures, with varied loop gain. Based on settling time vs. jitter curve fitting equation, obtained from PLL behavioral simulations, Section \ref{sec:proposed_fom} proposes a new Figure of Merit, modified to consider lock time also as PLL performance parameter. Section \ref{sec:fom_cmp} uses results from existing PLL designs \cite{ref:7cicc_10}-\cite{ref:2trans2_10}, to show that proposed Figure of Merit correctly rates PLL performance, based on overall improvement in specifications of lock time, jitter and power.

% existing Figure of Merit \cite{ref:fom_paper} is inadequate for designs, which achieve low jitter and power, at the cost of increased settling time.
%% Since Jitter is reduced at the cost of Lock Time, therefore it is proposed to have modified equation.
% Below section show trade-off analysis. Also, behavioral model simulation result with different DCO gains and different loop filter gain are shown. % ADPLL with loop-gain change mechanism, implemented in this work, is designed for fast settling. Therefore, direct trade-off between settling-time and o/p clock jitter is not observed with this architecture, as shown in Sec.\ref{sec:bbpll_tradeoff}.
%  So, behavioral model of TDC based ADPLL, having linear system response, is simulated in Sec.\ref{sec:tdc_tradeoff} with varied DCO gain, to analyze trade-off between settling time and o/p clock jitter.
% % \hfill mds
% %  
%  Section V compares Figure of Merit for existing PLL architectures.
% \hfill November 18, 2002
% \section{Existing Figure of Merit for PLL}
% \label{sec:old_fom}
% Existing Figure of Merit\cite{ref:fom_paper} used to benchmark PLL performance, does not includes effect of increase in settling time, while reducing loop gain to reduce jitter, as shown in Eqn.(\ref{eqn:fom}) \cite{ref:fom_paper} :-
% 
% \begin{equation}
% \label{eqn:fom}
% FoM = 10log\left[\left(\frac{\sigma_t}{1s}\right)^2\left(\frac{P}{1mW}\right)\right]
% \end{equation}\\
\section{Trade-off Analysis for Bang-Bang Phase Detector based PLL}
\label{sec:bbpll}
Bang-bang phase detector quantizes phase error between reference clock and
feedback clock to only two levels, similiar to signum function. Being a non-linear system, Bang-Bang Phase Detector based PLL has its response characteristics dependent on input phase/frequency error magnitude. Unlike PFD or TDC based architectures, wherein output response could be characterized with linear system transfer function in frequency domain; analysis for BBPLL has to be done in time domain. So, for analyzing trade-off between lock time and jitter in this system, settling time is derived using time-domain equations, by tracing trajectory of phase-detector output.
% , by tracing the trajectory of bang-bang phase detector output
% 
% Bang-Bang PLL, without gain change mechanism, is inherently slow settling loop; therefore, analysis in existing work concentrates on deriving loop parameters to optimize jitter, rather than settling time. Section \ref{sec:ts_bbpd} uses time-domain equations to trace trajectory of phase-detector output, for finding settling time equation; and thus, deriving non-linear relationship between settling time and jitter.\\
% (\cite{ref:bbpll_book}-\cite{ref:bbpfd_golden_ref}) 
\begin{figure}[h]
\label{fig:bbpll_arch}
\begin{center}
\includegraphics[scale = 0.67]{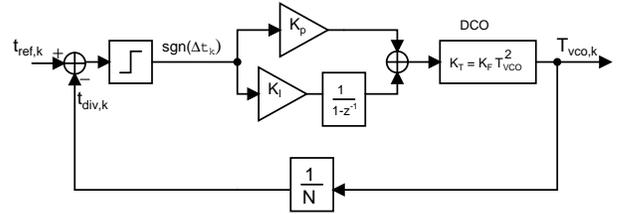}
\end{center}
\caption{{Bang-Bang Phase Detector based Digital PLL.}}
\end{figure}
\label{sec:bbpll_tradeoff}
\subsection{Settling Time derivation}
\label{sec:ts_bbpd}
Considering a case, wherein, at initial time-step (t = 0), phase error input is $\phi_o$ and frequency error input is 0; method explained below illustrates derivation of settling time for BBPLL, by tracing number of UP/DOWN pulses in loop's transient state.\\%, before DCO control word reaches required value

\begin{figure}[h]
\label{fig:bbpll_ts}
\begin{center}
 \includegraphics[scale = 0.43]{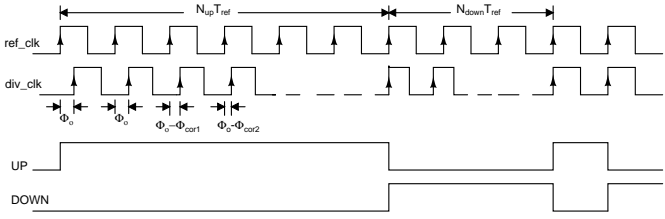}
\end{center}
\caption{ {Timing diagram of phase correction by Bang-Bang Phase Detector.}}
\end{figure}

\textbf{Step 1: Calculate number of UP pulses}\\ \\
(i) \textit{At t = $T_{ref}$} : Initial phase error = $\phi_o$\\ 
(ii)\textit{At t = $2T_{ref}$}: With loop correction, phase error decreases by increase in frequency :-
\begin{equation}
 \Delta{f}=(K_P + K_I)K_{VCO} \nonumber
\end{equation}

Phase correction, due to incremental frequency correction :
\begin{equation}
 \Delta{\phi}_{cor,1}=\frac{2\pi}{T_{ref}}\frac{(K_P+K_I)K_{VCO}}{f_o^2}N_{div} \nonumber
\end{equation}\\ 
(iii) \textit{At t = $3T_{ref}$}  : Phase correction, due to added frequency ($\Delta{f}=K_IK_{VCO}$), is :- 
\begin{equation}
\Delta{\phi}_{cor,2}=2\pi{K_I}\frac{K_{VCO}}{N_{div}}T_{ref} \nonumber
\end{equation}
(iv) \textit{At t = $N_{up}T$}    : Reference clock and feedback clock aligns in phase, if accumulated phase correction due to frequency correction done by loop in each reference cycle, becomes equal to initial phase error :-
\begin{IEEEeqnarray}{rCl}
\phi_o & = & 2\pi{T_{ref}}\left(\frac{K_{VCO}}{N_{div}}\right)\left[(K_P+K_I)N_{up}+K_I(N_{up}-1) \right. \nonumber\\
&& \left. \qquad \qquad \qquad \qquad + K_I(N_{up}-2)+... + K_I\right]\nonumber
\end{IEEEeqnarray}
\begin{equation}
\Rightarrow \phi_o = 2\pi{T_{ref}}\left(\frac{K_{VCO}}{N_{div}}\right)\left(N_{up}K_P+\frac{N_{up}(N_{up} + 1)K_I}{2}\right) \nonumber
\end{equation}\\
Solving above quadratic equation gives $N_{UP}$ value as,
\begin{IEEEeqnarray}{rCl}
\label{eqn:Nup}
\Rightarrow N_{up} & = &\frac{1}{2}\left[-\left(1+\frac{2K_P}{K_I}\right) \right. \nonumber\\  
&& \left. \pm\: \sqrt{\left(1+\frac{2K_P}{K_I}\right)^2 - \frac{4\phi_oN_{div}}{\pi T_{ref} K_P K_{VCO}}}\right]
\end{IEEEeqnarray}
Though, after $N_{up}$ cycles, phase of reference clock and feedback clock is aligned, but, resultant frequency error (normalized w.r.t $K_{VCO}$) accumulates to:-
\begin{IEEEeqnarray}{rCl}
 \triangle{f}' &=&\left[\left(K_P+\frac{N_{up}(N_{up}+1)}{2}K_I\right)-2K_{P}\right]\nonumber\\ 
&& =\: \frac{N_{up}.(N_{up}+1)}{2}K_I-K_{P} \nonumber
\end{IEEEeqnarray}
\textbf{Step 2: Calculate number of DOWN pulses}

Due to resultant frequency error ($\triangle{f}=f_{ref}-N_{div}f_{VCO}$) after $N_{up}$ cycles, phase error starts accumulating with time; which results in sequence of asserted DOWN cycles.\\

If phase error ($\phi_{err}$) becomes 0 after $N_{down}$ cycles, then, number of down pulses are calculated as :-
\begin{equation}
0 = 2\pi{\frac{K_{VCO}}{N_{div}}}T_{ref}\left[{\Delta{f'}N_{down}-\frac{N_{down}({N_{down}+1})}{2}K_I}\right]  \nonumber
\end{equation}
\begin{equation}
\label{eqn:Ndown}
\Rightarrow N_{down} = \frac{2\Delta{f'}}{K_I} - 1  = N_{up}(N_{up} + 1) - 2\frac{K_P}{K_I} - 1
\end{equation}
% (1) Phase accumulated due to initial frequency difference of $\triangle{f}_{init}$ error = $2\pi.\frac{K_{VCO}}{N_{DIV}}.\triangle{f}.T_{ref}.N_{DOWN}$\\
% 
% (2) Phase correction applied at each sampling instant = \\
% 
% (3) Phase correction afer $N_{DOWN}$ clock cycles :-
% 
% =$2\pi.{\frac{K_{VCO}}{N}}.T_{ref}.(N_{Down}.K_I + (N-1)_{DOWN.K_I}+ .......+K_I)$\\
Therefore, to derive settling time, equation for $N_{up}$ and $N _{down}$ cycles has to be recursively calculated until frequency error becomes 0, at phase error alignment :-
\begin{equation}
\label{eqn:ts_bbpll}
t_{settling} = (N_{up1}+N_{down1}+N_{up2}+.......)T_{ref}
\end{equation}

According to Eqn.(\ref{eqn:Nup})-(\ref{eqn:Ndown}), settling time decreases non-linearly with increase in filter proportional gain($K_P$), Digitally Controlled Oscillator gain ($K_{DCO}$), and ratio of filter's integral gain to proportional gain ($K_I/K_P$).
% Derived settling time is confirmed through simulation.
\subsection{Jitter Equation}
Peak to Peak Jitter for BBPLL, derived in \cite{ref:bbpfd_golden_ref} as Eqn.(\ref{eqn:bbpd_jitter}), indicates that jitter reduces with decrease in loop gain parameters, which are (i) $K_P$ (ii) $K_{DCO}$ (iii) $\frac{K_I}{K_P}$ ratio.
\begin{IEEEeqnarray}{rCl}
\label{eqn:bbpd_jitter}
\triangle{t_{pp}} &=& NK_PK_T\left[2(1+D)+(1+D)\frac{K_I}{K_P}\right. \nonumber\\
&& \left. \qquad  \qquad  +\: (1+D)^3\left({\frac{K_I}{K_P}}\right)^2  + O\left(\frac{K_I^3}{K_P^3}\right)\right]
\end{IEEEeqnarray}
\subsection{Settling Time vs Jitter Trade-Off}
\begin{figure}[h]
\begin{center}
\includegraphics[scale = 0.7]{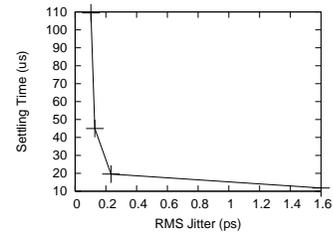}
\end{center}
\caption{ {Settling Time vs. Jitter trade-off for BBPLL.}}
\label{fig:bbpll_tradeoff}
\end{figure}

Eqn. (\ref{eqn:Nup})-(\ref{eqn:bbpd_jitter}) indicates that settling time and jitter holds inverse non-linear relation to filter parameters and Digitally Controlled Oscillator (DCO) sensitivity gain. For verifying this trade-off relation between settling time and jitter, Verilog-A model of BBPLL is simulated with different values of filter co-efficients. As shown in Fig. \ref{fig:bbpll_tradeoff}, settling time vs. jitter curve follows non-linear relationship, and that, jitter is reduced at the cost of increased settling time. This illustrates the need of adding lock time as a parameter to the commonly used FoM.
% 
% 
% \begin{equation}
% \label{eqn:fom_adpll}
% FoM = 10log\left[\left(\frac{\sigma_t}{1s}\right)^2\left(\frac{t_s}{1s}\right)^3\left(\frac{P}{1mW}\right)\right]
% \end{equation}
% 
% % Observed Jitter and Settling time matches with the simulation values.
% Relation is shown for Kp and ratio of Ki.Kp which has linear trade-off between settling time and trade-off
%*******************************************************************************
%------------------Trade-off for TDC based PLL----------------------------------
%*******************************************************************************
\section{Trade-off Analysis of TDC based PLL}
\label{sec:tdc_tradeoff}
For analyzing TDC based PLL with equivalent s-domain model (analogous to Charge Pump PLL), system transfer function is approximately given as :-\\
% analyzed in \cite{book:adpll} (\textit{without considering Normailzed DCO model}) :-
\begin{equation}
\label{eqn:adpll_tf}
H(s) = \frac{K_PK_{DCO}K_{TDC}s+K_IK_{DCO}K_{TDC}f_{ref}}{s^2+\frac{K_PK_{DCO}K_{TDC}}{N_{div}}s+\frac{K_IK_{DCO}K_{TDC}f_{ref}}{N_{div}}} \nonumber
\end{equation}
\begin{figure}[h]
\begin{center}
\includegraphics[scale = 0.65]{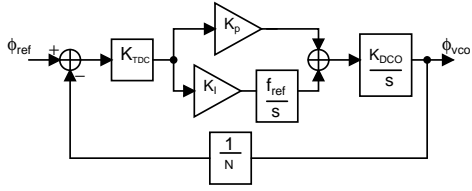}
\end{center}
\caption{{Linear s-domain model of TDC based PLL.}}
\label{fig:tdc_linear}
\end{figure}
\subsection{Settling Time}
Considering decay constant from system transfer function, settling time for Type-2 PLL is approximately given as :-\\
\begin{equation}
\label{eqn:ts_adpll}
  t_s = \frac{5}{\xi\omega_n} \propto \frac{10}{K_{TDC}K_PK_{DCO}} 
\end{equation}
Eqn. (\ref{eqn:ts_adpll}) indicates that with decreasing DCO gain and filter proportional gain, settling time increases linearly.
\subsection{Jitter Equation}
% In TDC based ADPLL, linear gain of TDC block is high, for high resolution delay elements. Therefore, loop filter proportional gain can be kept low ($K_P = 1$),so that DCO Control Word oscillates within 1 LSB count, to generate fractional control word.
When TDC-based PLL locks to frequency corresponding to fractional value input to DCO, control word oscillates between $\pm{K_P}{K_{DCO}}$. So, with lower DCO gain ($K_{DCO}$) and filter proportional gain ($K_P$), jitter reduces linearly for output frequencies corresponding to fractional control word :-
\begin{equation}
\label{eqn:jit_adpll}
  \Delta{t} = \frac{\Delta{f}}{f_o^2} \propto \frac{K_PK_{DCO}}{f_o^2}
\end{equation}
% Since, TDC-based ADPLL Settling Time vs. jitter is found by varying DCO Gain. In case of Bang-Bang based ADPLL, Loop Gain cannot be reduced beyond 4 to avoid oscillation in Loop Gain Change mechanism. 
% But, in case of TDC based ADPLL. since gain of TDC block is very high (=2000), therefore loop gain of filter can be kept low; 
\subsection{Settling Time vs Jitter Trade-Off}
\begin{figure}[h]
\begin{center}
\includegraphics[scale = 0.7]{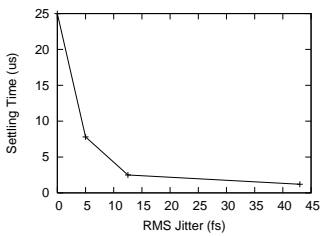}
\end{center}
\caption{ {Settling Time vs. Jitter trade-off for TDC based PLL.}}
\label{fig:tdc_tradeoff}
\end{figure}

Behavioral simulation of TDC based PLL model with varying DCO gain, indicates linear trade-off between settling-time and jitter (i.e. jitter increases and lock time decreases by same percentage with varied $K_{DCO}$) following linear curve fitting equation $\sigma t_s$, as shown in Fig.\ref{fig:tdc_tradeoff}. 
% 
% \textit{(Since, the periodic jitter is only due to quantization error of DCO and TDC, therefore, resultant jitter value in this behavioral model is very low.)}

So, based on Eqn.[\ref{eqn:ts_adpll}]-[\ref{eqn:jit_adpll}] and simulation results, Figure of Merit is proposed to be modified to include linear trade-off between settling time and jitter, by considering lock time in same ratio as jitter. %, as shown in Eqn. (\ref{eqn:fom_adpll}).
% , as shown in Eqn.[\ref{eqn:fom_adpll}] :-
%*************************************************************************
%----------------------Hybrid PLL with Adaptive Gain---------------------
%*************************************************************************
\section{Trade-off analysis for Hybrid PLL with adaptive loop gain}
\label{sec:hybrid_pll}
\begin{figure}[h]
\begin{center}
 \includegraphics[scale=0.65]{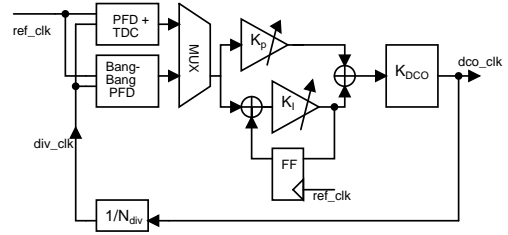}
\end{center}
\caption{ {Hybrid PLL with gain-change mechanism.}}
\label{fig:adaptive_gain_pll}
\end{figure}

Hybrid PLL, in Fig. \ref{fig:adaptive_gain_pll}, employs two different phase detector blocks for fast linear system response in transient state, and low jitter in settled state. Here, linear Phase-Frequency Detector (PFD) is activated in case of large input phase error, for fast coarse settling; and binary phase detector is activated for small phase error, to avoid the dead zone issue. Adaptive gain feature allows filter gain to be changed, according to input phase error magnitude.
\subsection{Effect of varying DCO Gain}
\begin{figure}
\centerline{\subfigure[]{\includegraphics[scale = 0.7]{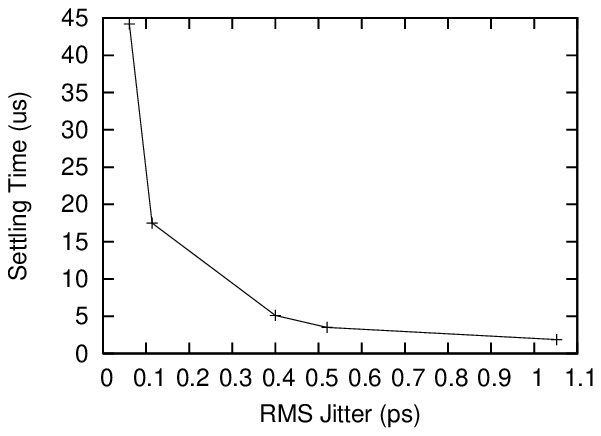}
\label{fig:a1_pll_dco_tradeoff}}
\hfil
\subfigure[]{\includegraphics[scale = 0.7]{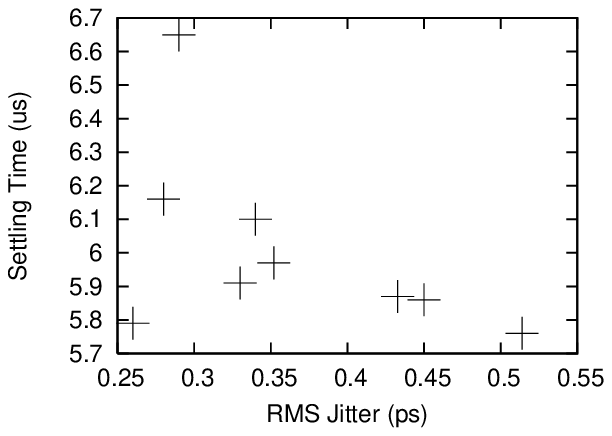}
\label{fig:a1_pll_filt_tradeoff}}}
\caption{Settling Time vs. Jitter trade-off for Hybrid PLL (a) with varying DCO gain (b) with varying filter gain in binary phase-detection mode.}
\label{fig_sim}
\end{figure}

During loop transient in Hybrid PLL, linear PFD is active as error-detection block for major portion of settling time. This results in loop settling curve to follow linear response, as per Eqn. (\ref{eqn:adpll_tf}) (which represents system response for linear PLL system). So, in case of Hybrid PLL also, linear trade-off between settling time and jitter is observed  in behavioral simulation with varied $K_{DCO}$, as shown in Fig. \ref{fig:a1_pll_dco_tradeoff}.
\subsection{Effect of varying filter constants, in binary phase detection mode}
In Hybrid PLL, non-linear bang-bang phase detector is activated only when loop reaches near to locked state. Therefore, reducing loop filter gain in bang-bang phase detection mode, can reduce jitter without significant linear increase in settling time. This overall performance improvement is verified by simulating Hybrid PLL Verilog-A model, with varied filter coefficients in binary phase detection mode. Simulation results in Fig. \ref{fig:a1_pll_filt_tradeoff} shows that jitter reduces with reduced filter gain (while binary phase detector is active in settled state), without significantly increasing settling time (since, lock time is mainly governed by loop parameters while PFD is active). So, this kind of PLL design, wherein all the specifications (Lock Time/Jitter/Power) could be simultaneously improved, can be benchmarked as having higher Figure of Merit.

% &&&&&&&&&&&&&&&&&&&&&&&&&7 Check &&&&&&&&&&&&&&&&&&&&&&&&&&&&&77
% Behavioral simulation of Bang-Bang DPLL with adaptive tracking, for varied loop filter gain, shows non-linear trade-off relation between settling time and o/p clock jitter, as shown in Fig.\ref{fig:bbpll_tradeoff}. Since, with loop gain change mechanism, coarse settling is done by linear PFD model, therefore, even by varying loop gain while BBPFD block is active, settling time does not alters as much as jitter changes.
% 
% 
% This non-linear trade-off between settling time and jitter is expected, from Eqn.(\ref{eqn:Nup}) and Eqn.(\ref{eqn:bbpd_jitter}). Therefore, with Verilog-A model simulation results, non-linear curve fitting equation for trade-off relation is derived as ${\sigma_t}{t_s}^{1.5}$.
% PFD based PLL with variation shown in KDCO shows linear relation In case of Hybrid PLL, settling time is mainly reached with linear part and follows linear response.With DCO gain variation, settling time is expected to linearly vary, because for most of the transient time, loop response is governed by linear PFD block.
%*************************************************************************
%----------------------Proposed Figure of Merit---------------------
%*************************************************************************
\section{Proposed Figure of Merit}
\label{sec:proposed_fom}
PLL performance analysis for different PLL architectures, in Sec. \ref{sec:bbpll}-\ref{sec:hybrid_pll}, shows inverse relation that settling time and jitter holds with loop gain. Considering lock time vs. jitter performance observed in linear PLL system, it is proposed to include settling time in Figure of Merit equation, in same ratio as jitter, as shown in Eqn. (\ref{eqn:fom_adpll}).
% Power dissipation of overall logic could be considered nominal in comparison to Analog block
\begin{equation}
\label{eqn:fom_adpll}
FoM = 10log\left[\left(\frac{\sigma_t}{1s}\right)^2\left(\frac{t_s}{1s}\right)^2\left(\frac{P}{1mW}\right)\right]
\end{equation}

%*************************************************************************
%-------------------------------PLL Performance Comparison----------------
%*************************************************************************
\section{PLL Performance Comparison}
\label{sec:fom_cmp}
Table-\ref{tab:result_cmp} uses results of existing PLL architectures (designed in 180nm-90nm technology, with output frequency in 0.5GHz-1.8GHz range), to compare performance benchmarking done with commonly used Figure of Merit versus proposed Figure of Merit in this work.

In Table-\ref{tab:result_cmp}, performance comparison for 130nm/90nm designs, indicates that for PLL \cite{ref:7cicc_10}\cite{ref:tj_wban_rfic_12} achieving lowest jitter at the cost of large settling time, existing FoM\cite{ref:fom_paper} proves as an inadequate benchmark by ignoring lock time performance degradation. Similarly, performance comparison for 180nm designs, indicates that PLL\cite{ref:2trans2_10} having lowest settling time at the cost of increased jitter is benchmarked for having lowest performance with FoM defined in \cite{ref:fom_paper} (which does not considers reduced lock time as improved PLL performance parameter).

Proposed Figure of Merit, on the other hand, correctly marks PLL designs for their overall performance, by considering inverse relation existing between settling time and jitter (as indicated in Table-\ref{tab:result_cmp}). 
% These observations signifies the need of adding settling time as performance parameter in PLL Figure of Merit equation, as shown in Eqn.(\ref{eqn:fom_adpll}).
% In , significance of proposed Figure of Merit is visible for PLL design in \cite{ref:7cicc_10}, which achieves low jitter and power, at the cost of very high settling time. If existing FoM\cite{ref:fom_paper} is used for benchmarking PLL performance, design in \cite{ref:7cicc_10} would be considered to give optimum performance; but, proposed FoM clearly shows that all PLL specifications are not equally optimized in this architecture.
\begin{table}[h]
\begin{center}
     \caption{Perfomance Benchmarking for PLL Architectures}
      \label{tab:result_cmp}
\renewcommand{\arraystretch}{1.4}
  \begin{tabular}{|c|c|c|c|c|c|c|c| }
\hline
Ref.&Tech. &Freq.& Jitter & Power & FOM                 & Lock & Proposed \\
    &     &       &        &       & of                & Time & FOM \\ 
    & (nm)&(GHz) &(ps)    &(mW)   & \cite{ref:fom_paper}&($\mu$s)  &(Eqn.\ref{eqn:fom_adpll})\\ \hline
\cite{ref:7cicc_10}              &130&1.56&0.38 &2.8 &-243.9&300 &-314.4\\ \hline
\cite{ref:3JSSC_10}              &130&1.35&3.7  &16.5&-216.5&3.84&-324\\ \hline \hline 
\cite{ref:tj_wban_rfic_12}       &90 &1.73& 4.15&1.13&-227.1&40  &-315       \\ \hline
\cite{ref:tj_vlsi_dat_12}        &90 &0.64&  4.9&3.8 &-220.4&4.67&-327\\ \hline
\cite{ref:tj_iscase_10_delayline}&90 &0.48& 5.8&3   &-219.9&2   &-333\\ \hline \hline
%=====================180nm Shown in FoM====================================
\cite{ref:1trans2_11}           &180&1.2&3.5&18&-216.6&5&-322\\ \hline
\cite{ref:10vlsi_11}            &180&1.56&9.7&16.2&-208.2&26&-299\\ \hline
\cite{ref:2trans2_10}           &180&0.446&70&14.5&-191.5&0.5&-317.5\\ \hline
%====================Lower Technology Nodes=================================
% \cite{ref:6trans2_10}&1&65nm&4.5&8.4&-213.1&17&-308\\ \hline
% \cite{ref:6cicc_11}&1.8&55nm&0.138&41.6&-241&23&-333.7\\ \hline
% \cite{ref:1isscc_10h}&3.2&22nm&0.8&3.4&-236.62&2.3&-349.38\\ \hline
% % \cite{ref:tdc_cmp}&4.08GHz&90nm&0.682&9.6&-233.5&0.74&-356 \\ \hline
% % \cite{ref:5cicc_11}&3.96GHz&90nm&0.78&9.6&-233.34&0.63&-356.34\\ \hline
% \cite{ref:2isscc_12}&1.5&32nm&1.45&2.5&-232.79&25&-324.83\\ \hline
%============================================================================
%========================10 GHz 90nm========================================
% \cite{ref:1isscc_09}&10&90nm&0.9&7.1&-232.4&6.89&-335\\ \hline
%============================================================================
% [6]&0.95GHz&65nm&3.1&10&-220&12.8&-318\\ \hline
% [7]&0.55GHz&130nm&56.36&1.25&-204&4.7&-310.56\\ \hline
%===========================================================================
\end{tabular}
\end{center}
\end{table} 

\begin{figure}[h]
\begin{center}
\includegraphics[scale = 0.87]{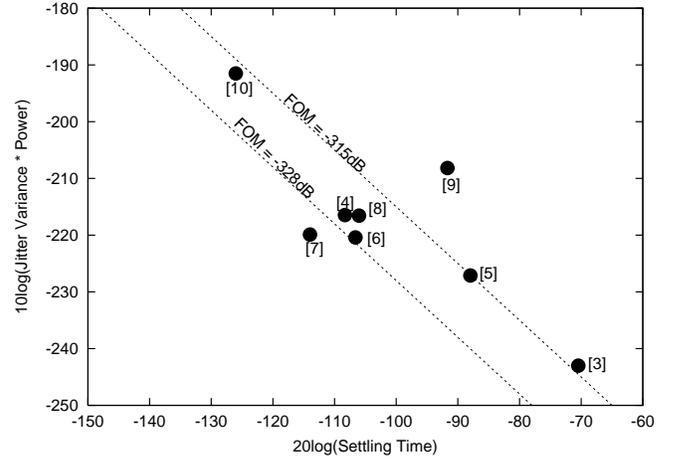}

\end{center}
\caption{PLL performance benchmarking with proposed FoM.}
\label{fig:new_fom}
\end{figure}

Fig. \ref{fig:new_fom} shows performance of existing PLL designs, with consideration to lock time, along with output jitter and power dissipated by PLL system. 
\section{Conclusion}
In this work, with analysis and behavioral simulation of different PLL architectures, trade-off between settling time and jitter is shown. Since, most PLL designs are able to achieve low jitter only at the cost of increased settling time, therefore, it is proposed to include lock time also as performance parameter in Figure of Merit for PLL.

Based on analysis of linear PLL system, we have proposed a Figure of Merit for PLL, wherein settling time is considered in equal proportion as jitter, for deciding overall PLL performance.
% Analogous equation shown for Charge-Pump PLL
% Above FoM is valid for any kind of linear PLL.

% that's all folks
\end{document}